\documentclass[preprint]{aastex}
\def\ltsima{$\; \buildrel < \over \sim \;$}
\def\lsim{\lower.5ex\hbox{\ltsima}}
\def\gtsima{$\; \buildrel > \over \sim \;$}
\def\gsim{\lower.5ex\hbox{\gtsima}}

\slugcomment{Submitted to the Astronomical Journal}
\shorttitle{First QSO lens(es) from VST-ATLAS and WISE}
\shortauthors{Schechter et al.}

\begin{document}
\title{First lensed quasar systems
from the VST-ATLAS survey: one quad,
two doubles and two pairs of lensless twins}

\author{Paul L. Schechter\altaffilmark{1}}
\affil{MIT Kavli Institute for Astrophysics and Space Research,
Cambridge, MA 02139}

\author{Nicholas D. Morgan}
\affil{Staples High School, Westport, CT}

\author{B. Chehade}
\affil{Durham University} 

\author{N. Metcalfe}
\affil{Durham University} 

\author{T. Shanks}
\affil{Durham University} 

\author{Michael McDonald\altaffilmark{1}}
\affil{MIT Kavli Institute for Astrophysics and Space Research,
Cambridge, MA 02139}

\altaffiltext{*}
{This paper includes data gathered with the 6.5 meter Magellan 
telescopes located at Las Campanas Observatory, Chile.}
\altaffiltext{1}{MIT Department of Physics}

\begin{abstract}

  We have analyzed images from the VST ATLAS survey to identify
  candidate gravitationally lensed quasar systems in a sample of WISE
  sources with $W1 - W2 > 0.7$.  Results from followup spectroscopy
  with the Baade 6.5 m telscope are presented for eight systems.
  One of these is a quadruply lensed quasar,  and two are doubly
  lensed systems.  Two are projected superpositions of two quasars
  at different redshifts.  In one system two quasars, though at the
  same redshift, have very different emission line profiles, and
  constitute a physical binary.  In two systems the component
  spectra are consistent with the lensing hypothesis, after allowing
  for micro-lensing.  But as no lensing galaxy is detected in these
  two, we classify them as {\it lensless twins}.
  More extensive observations are needed to establish whether they are
  in fact lensed quasars or physical binaries.
  
\end{abstract}

\section{Introduction}

Doubly and quadruply lensed quasar systems are valuable for widely
disparate purposes.  Treu and Marshall (2016) present a current survey
of the use of time delay measurements for cosmography.  The
micro-lensing of lensed quasars can be used to determine sizes for the
emitting regions of quasars (Rauch and Blandford 1991; Agol and Krolik
1999; Pooley et al, 2007; Blackburne et al 2011) and to measure the
dark matter fraction in lensing galaxies (Schechter and Wambsganss
2004, Pooley et al 2012, Jimenez Vicente et al 2015).  For each of these
efforts the accuracy achieved is limited by the relatively small
number of lensed systems.

The most productive lensed quasar discovery program to date has been
the Sloan Digital Sky Survey Quasar Lens Search, henceforth SQLS
(Inada et al. 2012), which yielded a statistical sample of 26
lensed quasar systems brighter than a limiting magnitude $i_{lim} =19.1$
over 8000 square degrees in the redshift range $0.6 < z < 2.2$.
An additional 36 systems that did not satisfy all of the selection
criteria were also catalogued.  Of 62 systems {\it in toto}, 40
were newly identified.

The ATLAS survey, carried out with VLT Survey Telecope (Shanks et
al. 2015), promises to yield comparable if not greater numbers of
lensed quasar systems.  Its $ugriz$ limiting magnitudes are nearly
identical to those of SDSS.  While its $ugriz$ photometry covers only
4500 square degrees, the typical ATLAS seeing is 3/4 that of SDSS
(Shanks et al. 2015), permitting the discovery of quasar pairs with
smaller separations.

We have undertaken a search for lensed quasars in the ATLAS survey,
and report here the first newly discovered lensed quasars:  a quadruple
system and two doubly lensed quasars.  We also report two
systems that have two nearly identical quasar spectra but for which no
lensing galaxy has been detected.

In \S2 we outline our method for identifying candidate lensed quasar
systems, using the WISE catalog and ATLAS $ugriz$ cutouts to choose
candidates for spectroscopic and direct imaging followup.  The method
will be described in greater detail in a forthcoming paper.  In \S3 we
describe direct imaging and spectroscopic followup observations of
seven cadidate systems obtained with IMACS on the Baade 6.5 m
telescope of the Magellan Observatory.  In \S4 we analyze these
observations.  In \S5 we present simple models for the the newly
discovered quadruple system, WISE 2344-3056  and for one of the
  doubles, WISE 2304-2214.  We discuss our method and results in \S6
and summarize our findings in \S7.

\section{Selection of Candidate Lensed Quasar Systems}

\subsection{Colors for marginally resolved lensed quasar systems}

The colors derived for an object from a survey like ATLAS involve a
number of implicit assumptions.  Bright objects that are deemed point
sources are used to determine a point spread function (PSF)
appropriate to the exposure, usually dominated by the atmospheric
seeing.  Sources that appear extended with respect to this PSF might
be fit with a Sersic (1963) profile.  For sources that do not
appear to be extended, one uses the adopted PSF to calculate a
magnitude.

A gravitationally lensed quasar system is a composite object
consisting of multiple images of a quasar and one or more lensing
galaxies.  Typical image separations are $\sim 1\arcsec$, to within a
factor of two.  It is unfortunate that the typical seeing in a survey
like ATLAS is also of order one arcsecond.  Were it very much better,
one would detect the components of the lensed quasar as distinct
objects and compute colors for each.  Were it very much worse, one
might treat the system as a point source.\footnote{One might in
  principle convolve survey images with a broadening function, perhaps
  blurring those taken in different filters to a common PSF, and
  produce a catalog for the smeared data, but only with the
  expenditure of considerable resources.}

When the resolution of a survey is {\it comparable} to the image separation,
the magnitudes derived for composite systems suffer systematic errors
from the mismatch between the assumed PSF and the object.
Moreover these systematic errors will vary with the seeing.  While one
might hope to select lensed quasars using their catalogued colors,
systematic errors may cause systems to be missed.  

The resolution of the WISE survey (Wright et al. 2010) is
substantially worse than that of ATLAS, and WISE magnitudes for all
but the widest lensed quasar systems do not suffer from marginal
resolution.  Our approach is therefore to make a first level selection
based only on WISE colors.  This produces a candidate list
sufficiently small that one can then retrieve the ATLAS survey images
for each remaining object and determine whether they are consistent
with its being a system of two or more sources.  Magnitudes are then
computed by adopting a fixed configuration, with the same number of
components, for all filters, thereby mitigating the systematic errors
associated with marginally resolved systems.

\subsection{W1 - W2 color selection}

Stern et al (2012) have shown that the WISE $W1-W2$ color
($m_{3.6\mu}- m_{4.5}\mu$) can be used to isolate quasars from stars,
and, to a lesser extent, from galaxies.  The underlying explanation is
that the optical light from a quasar is thermal emission from an
extended region with a range of temperatures, in which case the red
tail of the distribution is redder than a blackbody Rayleigh-Jeans
spectrum, which has $W1-W2 \approx 0$ in the Vega system.
We adopted a criterion, $W1 - W2 > 0.70$, that struck a
balance between including lensed quasar systems and isolating them
from other objects.  Eighty percent of the confirmed SQLS lensed
quasar systems satisfy this criterion.  Lensing galaxies have bluer
$W1 - W2$ colors than quasars.  Systems in which the light from the
lens dominates that of the quasars will not be included in our sample.
The fraction of SQLS quasar systems that are bluer than our criterion
is larger for the fainter systems.  To keep the number of candidates
manageable, we limited our sample
to objects with $W1 < 15$ or $W2 < 14.45$.

In the south galactic cap, ATLAS covers a region with $21^h30^m <$ RA
$< 4^h00^m$ and $-40^\circ <$ Dec $< -10^\circ$ in the $ugriz$ filters
with $0\farcs213$ pixels.  (Shanks et al 2015).  In the north it
covers a region with $10^h00 <$ RA $< 15^h30^m$ and $-20^\circ <$ Dec
$< -2.5^\circ$.\footnote{A second northern region with $10^h00^m <$ RA
  $< 15^h00^m$ and $-35^\circ <$ Dec $< -20^\circ$ is being surveyed
  in $i$ and $z$, with $u$, $g$, and $r$ being observed in a separate
  ESO program (095.A-0561, PI L. Infante), but was not included in the
  present search.}
The completed survey will include 4292
fields, each one degree square, located on $0\fdg98$ centers. As of
mid-March 2016 there were 3650 fields for which data in $u$,$g$ and
two or more of $r$,$i$ and $z$ were available.

ATLAS survey data in at least one filter was available, as of
mid-March 2016, for 144,700 unique sources that satisfied our WISE
criteria.

\subsection {Cutouts}

For each source in the overlap between WISE and ATLAS we downloaded
$12\arcsec$ square $ugriz$ FITS subrasters (58 pixels on a side) from
the University of Edinburgh Wide Field Astronomy Unit's OmegaCam
Science Archive (henceforth OSA; Hambly et al 2008; Cross et al 2012).
The size of these ``cutouts'' was chosen to be larger than most known
quasar/galaxy lens systems while minimizing the possibility of
including an unrelated source in the field.  

\subsection{Splitting blended images}

For each source the available cutouts were analzyed with a
  program that incorporates many of the core subroutines from the
{\tt DoPHOT} photometry program (Schechter, Mateo and Saha 1993).
{\tt DoPHOT} uses an elliptical profile that approximates a Gaussian
near the core but has broader wings.

In each of the three filters with the best seeing, as recorded in the
FITS header of the cutout, we attempt to split candidates into two
sources with a common quasi-Gaussian shape.  From these we chose the
two source fit that shows the greatest improvement over a single
extended object (as measured by our goodness of fit parameter) as our
``anchor" estimate of the separation between components.

For as many of the $ugriz$ filters as we have cutouts, we carry out
two source fits, with the separation constrained to our anchor value,
but allowing the fluxes to vary along with a common set of shape
parameters and an overall position.  We call these constrained separation
fits.

If the flux ratio from one of these constrained separation fits is
very different from the anchor value (3 magnitudes) at a high level of
significance ($10 \sigma$) we take the anchor splitting to be
spurious.  This frequently happens at smaller separations, at which
trailed or astigmatic images can cause the object to look elongated or
double in one exposure.

\subsection{Weeding out galaxies}

In addition to quasars, our WISE quasar-colored systems include star
forming galaxies, which often come in close pairs.  Moreover, with a
thousand unlensed quasars for every lensed quasar, we expect
accidental projections of foreground galaxies close to quasars.

Every constrained separation fit gives us an elliptical footprint that
can be compared with the seeing as recorded in the FITS header for
that filter.  Pairs of galaxies are expected to have larger
footprints, as measured by the area of the quasi-Gaussian, than the
stellar PSF.  After some experimentation, we decided to eliminate as a
probable galaxy pair any system for which the {\it area} of the
constrained separation footprint is larger than that of the stellar
PSF by 2 pixels in all three of the filters with the best seeing.

We carry out a second test using the cutout that yielded our anchor
splitting, fitting for two sources whose shapes are not constrained to
the same elliptical Gaussian.  This adds three additonal shape
parameters, but the fits converge for roughly 65\% of our candidates.

We exclude systems for which the minor axes of the two components
differ by more than than a factor of $\sqrt{2}$.  We reason that while
our splitting might lump together a close pair of quasar images, or a
quasar image and the lensing galaxy, the minor axis of the fitted
ellipse ought not to be much larger than the PSF of a star.  But
instead of using a nominal PSF, we use use the fit to the other split
component.  We further ighten this criterion if the major axis of the
wider component is very much larger than the narrower one, excluding
systems for which the minor axes differ by more than a factor of
$2^{1/4}$ and for which the ratio of the areas is larger than a factor
of two.
 
\subsection{Ranking on the basis of achromaticity}

The word ``achromatic'' is often used to describe gravitational
lensing.  While color may vary from one part of an extended source to
another, if the source is small compared to the Einstein ring of a
gravitational lens, the multiple images will all have the same color.

This is less powerful than one might hope in discriminating between
lensed quasar systems and chance superpositions of objects for two
reasons.  First, lensed systems also include lensing galaxies, the
light from which will be split disproportionately between the two
images.  Second, especially at brighter apparent magnitudes, the
quasars are micro-lensed by the stars within the lensing galaxies and
are somewhat extended compared to the micro-lensing Einstein rings.
This leads to differential micro-lensing (Blackburne et al 2011).

We use the constrained separation flux ratios at each observed
wavelength, expressed in magnitudes, $\delta m_\lambda$, to fit for a
slope, $d \delta m_\lambda/d \log \lambda$.  We adopt a rough guess
of the mean slope for a lensed system $<\alpha> = 0.217$, and of the
scatter in that slope, $\sigma_\alpha = 0.325$, and score systems
based on their deviation from the mean slope.\footnote{The mean slope
  is non-zero because fainter images
  are more likely to include more of the red light from the lensing
  galaxy.}  We also score systems on the absence of scatter from the
observed slope and finally score systems on the consistency of the $u$
filter flux ratio with those in the other filters.  A final ranking is
computed by taking the geometric mean of these three scores.  Details of our
scoring system (which we continue to refine as we observe more
candidates) will be presented in a forthcoming paper.

\subsection{Optical colors}

We have argued above that catalogued optical colors may be unreliable
for lensed quasar systems because catalog photometry explicitly or
implicitly assumes a light distribution over the detector pixels
that is inappropriate for a lensed quasar system.  Our constrained
separation model should produce better (but hardly perfect) colors.

For this, our first pass through the data, we restricted ourselves to
the simplest of quasar color criteria, ultraviolet excess (henceforth
UVX).  This works well for quasars with $z < 2.2 $ (Richards et al
2001) but would exclude the roughly 25\% of unlensed quasars in an
SDSS-like survey that have higher redshifts.

Optical colors were computed adding the two fluxes from the 
constrained separation fits in each of the filters.  The
photometric zeropoints embedded in the OmegaCam FITS headers were used
to create $u - g$ colors in a Vega-like $ugriz$ system (Shanks et al
2015).  We ultimately adopted $u - g < -0.5$
as our UVX criterion,
which seems to exclude narrow emission line galaxies and white dwarf
pairs, but to include lensed quasars at $z < 2.2$.

Applying the recipe described in this section, we are left with a
ranked list of candidate lensed quasar systems.

\section{Spectroscopic and Direct Observations and Reductions}

From June 2015 through April 2016,  and again in November 2016
spectroscopic observations were
carried out for roughly a dozen highly ranked systems with both the
f/2 and f/4 cameras of the Inamori Magellan Areal Camera and
Spectrometer, henceforth IMACS (Dressler et al 2010) on the Baade
6.5-m telescope of the Magellan telescopes.  In Table 1 we give
coordinates for eight of the objects observed, their rankings and
colors, and descriptions of the resulting spectra.

\subsection{Choice of Systems}

While guided by rank, the actual choice of systems to observe also
depended upon seeing, as some of the systems are quite close, and upon
cloud cover.  Pairs of blue stars and narrow line galaxies with $ -0.5
< u - g < 0$ that predominated in our first observations led us to
tighten our UVX criterion to $ u - g < -0.5$ in subsequent runs.

\subsection{Direct Imaging of WISE 2344-3056}

The system WISE 2344-3056 was given top priority for observation in
December 2015 because its appearance in the ATLAS images suggested
a quadruple system.
Figure 1 shows the VST ATLAS image of WISE 2344-3056 in the $g$
filter, which gave the best splitting of the object.  We have
superposed the elliptical FWHM contours from the anchor fit
to this image.

The minor axis of the anchor fit is $1\farcs11$,
slightly less than the $1\farcs18$ seeing reported in the image
header. The major axis is $1\farcs41$ and elongated so that each
ellipse includes two of the four images.

The system only barely survived being eliminated as a pair of
galaxies, suggesting that we may need to relax the constraint on the area
of the anchor footprint described in \S2.5.

The spectrocopic mode used with IMACS required the taking of one or
more short direct images to position the object in the slit.  The
first of those obtained for WISE 2344-3056, in Sloan $r$, confirmed
the suspicion that it was at least triple so two more images were
taken in Sloan $i$.  The left panel of Figure 2 shows one of the $i$
images, taken in $0\farcs55$ seeing with the f/2 camera.

The debiased and flatfielded frames were analyzed using the program
{\tt DoPHOT} (Schechter et al 1993).  The standard version of the
program found all four quasar images on the $r$ frame and needed only
minor nudging to find all four on the other two.  The right panel of
Figure 2 shows the residuals from one of those fits, using the point
spread function of a nearby star as the empirical PSF.  While
the residuals show little or no trace of a lensing galaxy, the
figure is somewhat deceptive.  If one allows for a {\it fifth} point
source with its position fixed at the expected position of the lens (see
\S5 below), the magnitudes of the four quasar images decrease by
hundredths of a magnitude, the positions spread out radially by
roughly 0.1 pixel each, and the flux from the hypothesized fifth
image is $1.5 \pm 0.5 $ magnitudes less that of image D.

Astrometry was carried out on one of the $i$ frames using 
catalogued positions from the ATLAS survey, with rms residuals
of $0\farcs 1$.  Results are given in Table 2.  The four positions
are indicated by the blue circles in Figure 1.

Fluxes relative to the brightest image were likewise computed using
{\tt DoPHOT}.  These were then used to compute magnitudes assuming a
combined Petrosian $i$ magnitude of 19.15 as given in the ATLAS
catalogue, which reports magnitudes in an AB system (Shanks et al
2015) rather than in the Vega-like system of the FITS headers.  To the
extent that the quasar has varied in the 4 years between the ATLAS
and IMACS exposures, these will share a common systematic error.


\subsection{Direct Imaging of WISE 0326-3122}

A 30 s direct image of WISE 0326-3056 in the Sloan $r$ filter in
$0\farcs64$ seeing was obtained with the IMACS f/2 in 
  acquiring the object for
spectroscopy.  The debiased and flatfielded frame was analyzed with
{\tt DoPHOT}.  Figure 3 shows the original image of the candidate
system and the same image with best fitting PSFs subtracted.  The
residuals show little or no trace of a lensing galaxy.
In contrast to the case of WISE 2344-3056, the residuals are not
deceptive.  If one allows for a {\it third} point
source with its position fixed near he expected position of the lens
(one third of the way from the fainter image to the brighter image)
one gets a {\it negative} flux with an amplitude only 1\% that of the
fainter image.  This limit is sensitive to the assumed position
for the lens, which as we see in the case of WISE 2304-2214 below,
can be quite different from the expectation.

\bigskip
\subsection{Direct Imaging of WISE 2304-2214}

In setting up for spectroscopy of WISE 2304-2214 a 30 s acquisition
image in the Sloan $r$ filter was obtained in $0\farcs63$ seeing with
the IMACS f/4 camera binned $2\times2$.  There appeared to be a
lensing galaxy in between two pointlike images so a 60s image in the
Sloan $i$ filter was obtained following spectroscopy.  Both exposures
were analyzed by fitting two scaled versions of a stellar template and
a quasi-Gaussian.  The separations determined from the $i$ exposure
were enforced on the $r$ exposure.

Figure 4 shows the original Sloan $i$ image, the same image with all
three sources subtracted, at 10 times the contrast, and again with
only the two point sources subtracted at 4 times the contrast.  We
take the central object to be the lensing galaxy.  The scale is
$0\farcs221$ per pixel.

In Table 3 we give positions and magnitudes for all three sources.
The latter are derived from the catalogued ATLAS magnitudes for the
stellar templates, two of which were used for each filter.  The
quasi-Gaussian fit yields shape paramters for the lensing galaxy.  We
deconvolve that quasi-Gaussian using a similar fit to the template
star and find the semi-major and semi-minor axes to be 0.58 and 0.25
pixels respectively, at position angle $-68\fdg6$, which is $9\fdg5$
off the pependicular to the line connecting the two images.  Similar
results were obtained from the $r$ exposure.  It is noteworthy that
contrary to the expectation for simple isothermal sphere models the
lensing galaxy is closer to the brighter image.  In this regard it is
similar to the case of HE1114-1805 discussed in \S4.1 below.

\subsection{Spectra}

Spectra for the objects in Table 1 were obtained using either the
``short'' f/2 camera or ``long'' f/4 camera on IMACS, in both cases
using a 3800-7000 \AA\ blocking filter. Dispersers with 300 lines
mm$^{-1}$ were used on both cameras: a grism blazed at $17\fdg5$ on
the short camera and a grating blazed at $4\fdg3$ on the long camera.
On the short camera the spectra were binned by 2 pixels in the
spectral direction (except where noted).  On the long short camera
they were binned by 4 pixels in the spectral direction and 2 pixels (with one
exception) along the slit.  The 0.9\arcsec\ slit was oriented to
obtain spectra of both components of the double systems.

The spectra were bias-subtracted and flattened using standard
procedures, and cosmic rays near the extraction paths were identified
by eye and replaced with interpolated values along each row of the
detector. Wavelength calibration was provided by Argon lamp lines
taken during the afternoons. A multi-order
polynomial fit as a function of both the spatial and dispersion
directions was used for the wavelength solution and gave typical fit
rms values with respect to the reference line list of 0.3 \AA\ or
better for all target chips.  The dispersion ranged from 2.2 to 2.6
\AA\ pixel$^{-1}$ on the short camera and from 2.8 to 3.0
\AA\ pixel$^{-1}$ on the long camera. Sky background was subtracted
using linear interpolation along each row of the detector. The spectra
show gaps near 6550 \AA\ on the short camera and near 5300 \AA\ on the
long camera due to the physical spacing between CCDs on the IMACS
cameras.

When the seeing permitted, spectra were extracted for the individual
components of each system. The was accomplished by fitting multiple
overlapping Gaussian profiles to each spatial row of the detector.  We
performed a preliminary fit for each row to obtain average component
separations, and then a final extraction where the overall position,
common FWHM, and component brightnesses were allowed to vary but the
relative separations were fixed at the average values.

Figures 5-7 show the extracted spectra for the objects in Table 1;
note that no flux calibration was performed. The displayed
spectra were top-hat smoothed using a 3 pixel window in the dispersion
direction for cosmetics. Gaussian profile fitting was performed on the
unsmoothed spectra and we plot the base-10 log of our fitted Gaussian
profile areas along the ordinate. We were able to extract individual
spectra for WISE 0145-1327, WISE 0326-3122, WISE 1051-1142, WISE
1427-0715, WISE 2215-3056, WISE 2304-2214 and WISE 2329-1258. For
these systems an inset at the lower right of each plot shows the Gaussian
decomposition of the two components for a single CCD row of the
detector.  The small separation of WISE 2344-3056 precluded decomposing
the spectra, so a single Gaussian was used to extract the combined spectrum.

\section{Interpretation of Spectra}

\subsection{Binary quasar, lensed quasar, or lensless twins?} 

The words ``binary quasar'' are used to describe two distinct quasars
at the same redshift, as opposed to two images of a single lensed
quasar (Hennawi et al 2006).  But when one observes a pair of quasars,
conclusive discrimination between these two alternatives is not always
straightforward (Wisotzki et al 1993; Kochanek et al 1999; Mortlock et
al 1999).

The history of HE1104-1805 is instructive in this regard.  Wisotzki
and collaborators (1993) found that the two components differed in the
slopes of their continuua and in the equivalent widths of their
emission lines, but that the shapes of their emission lines were
identical.  They argued that the spectral differences might be due to
micro-lensing by stars in the lensing galaxy.  But they detected no
lensing galaxy at $R \lsim 24$.  Wisotzki et al (1995) subsequently
observed correlated changes in the continuum flux of the two systems,
which they took as confirmation of the lensing hypothesis.  Courbin et
al (1998) conclusively detected the lensing galaxy, much closer to the
brighter image than might naively have been expected.  Crude
interpolation between filters in subsequent HST observations (Remy et al 1998)
would give $R \sim 22$ for the lensing galaxy.  Lidman et al (2000)
measured a lens redshift of $z = 0.729$.

The circumstances of two of our systems are similar to those of
HE1104-1805 in 1993.  The spectra differ, but no more than they might
under the micro-lensing hypothesis.  Still, we observe no
lensing galaxy.  We think it premature to call such systems binary
quasars, and instead refer to them as ``lensless twins.''  If lensing
galaxies or correlated variations are ultimately observed, they will
be classified as lens systems.

But if only upper limits can be measured for a lensing galaxy, careful
modeling is needed to establish that those upper limits are
inconsistent with plausible lensing scenarios.  Alternatively, higher
signal-to-noise spectra or spectra of narrow emission lines might show
significant differences in the line profiles or redshifts, ruling out
the lensed system hypothesis.

In the sequel to the SQLS, the SDSS-III BOSS Quasar Lens Survey, More et
al (2016) are similarly circumspect in not drawing strong conclusions
about lensless twins for which no lensing galaxy is
observed.

\subsection{WISE~0145-1327: a projected pair}

WISE~0145-1327 is a chance projection of two quasars.  The brighter of
the pair is at a redshift of $z=1.0902 \pm 0.0005$ from a Gaussian fit
to the MgII emission line.  The fainter object is at a higher redshift
of $z = 1.9749 \pm 0.0005$ based on its CIV broad emission line.
The spectra are shown in Figure~5.

\subsection{WISE~0326-3122:  lensless twins  at $z=1.34$}

The two quasar images of WISE~0326-3122 have nearly identical
redshifts and spectral flux ratios.  Both show a CIII] broad emission
line at around 4460 \AA.  Gaussian fits to the CIII] profiles yield
identical source redshifts of $z = 1.3342 \pm 0.0016$ for the
brighter object and $z=1.3336 \pm 0.0026$ for the fainter object.
This redshift places the MgII broad emission line inside the IMACS
chip gap with only a hint of the feature's wing visible for the
brighter component.  There is also a MgII
$\lambda\lambda$2796,2803 absorption doublet at $z = 0.5080 \pm
0.0001$ present in the brighter component's spectrum.  The flux
ratio between components is also remarkably constant at about
3.2:1 (0.5 dex in Figure~5) over the entire IMACS spectral range.
This makes it all the more surprising that no lensing galaxy is
observed in the PSF subtracted image (Figure~3).  We take the two
sources to be lensless twins.

\subsection{WISE~1051-1142: lensless twins at $z=0.88$}

The two quasar images of WISE~1051-1142 have nearly identical
redshifts.   The single prominent emission line just blueward of
the IMACS chip gap is likely MgII due to the absence of other
emission features in the observed wavelength range.  Gaussian fits
to the emission profiles yield source redshifts of $z = 0.8839 \pm
0.0002$ and $z = 0.8811 \pm 0.0008$ for the brighter and fainter
components, respectively, which overlap at the 3-4$\sigma$ level.
There are no obvious signs of intervening absorption.  The flux ratio
between components varies from 6:1  at the blue
end to 7:1  at the red end, not inconsistent with
wavelength-dependent continuum microlensing seen in other lensed
quasars.    We take the two sources to be lensless twins.

\subsection{WISE~1427-0715: a projected pair}

WISE~1427-0715 is a chance projection of two quasars.  The brighter
component shows the MgII broad emission line at about 6230 \AA\,
yielding a source redshift of $z = 1.2258 \pm 0.0004$.  The only
prominent emission line at 4820 \AA\ in the fainter component is
likely MgII due to the absence of other features in the observed
wavelength range, yielding a source redshift of $z = 0.7218 \pm
0.0004$. A MgII $\lambda\lambda$2796,2803 absorption doublet is also
seen in the brighter component's spectrum at a fitted redshift of $z =
0.7205 \pm 0.0004$ and thus associated with the host galaxy of the
fainter quasar.  The spectra are shown in Figure~6.

\subsection{WISE~2215-3056: a binary quasar at $z=1.34$}

WISE~2215-3056 is a binary quasar.  Its components have similar
redshifts but have notable spectral differences.  The brighter
component is at $z = 1.3474 \pm 0.0003$ based on its MgII broad
emission line profile and shows strong absorption features blueward of
its CIII] and MgII emission lines.  The fainter component has a source
  redshift of $z = 1.3508 \pm 0.0004$ based on its MgII profile, 
  a $> 5\sigma$ difference from the brighter component, but shows
  neither of the two absorption features seen in the brighter
  component.  There is also a prominent FeIII/UV 48 broad emission
  line of comparable intensity to CIII] in the brighter component that
    is lacking in the fainter companion.  Despite the similar
    redshifts, the spectral differences argue for two separate
    quasars.  The spectra are shown in Figure~6.

\bigskip    
\subsection{WISE~2304-2214: a doubly lensed quasar at $z=1.42$}

WISE~2304 -2214 is a doubly lensed quasar at z=1.423.  The quasar
spectra shown in Figure~6 have emission at CIII], CII] and MgII.
Gaussian fits
to the MgII emission profiles yield source redshifts of $z = 1.4222 \pm
0.0004$ and $z = 1.4218 \pm 0.0003$ for the brighter and fainter
components, respectively.
The brighter component
has an MgII absorption doublet at $z=0.6362 \pm 0.0001$ that is much weaker, if
not totally absent, in the fainter component.  Fitting a third
Gaussian component of the same width between the two quasar
components yields a spectrum of what appears to be the lensing
galaxy with Ca H\&K absorption near 5700\AA  giving $z_{gal} =
0.4455 \pm 0.0004$.

The emission line equivalent widths are substantially smaller in the
brighter component, suggesting that one or both components are
micro-lensed.  We make a crude estimate of emission line flux ratios
by comparing the peak flux in the emission line to the flux in the
adjacent continuum.  We get emission line flux ratios of $B/A = 0.99$
for the C III] line and and $B/A = 1.06$ for the Mg II line.
  
\subsection{WISE~2329-1258: a doubly lensed quasar at $z=1.31$}

The two quasar images of WISE~2329-1258 have nearly identical
redshifts with a rich set of absorption features present in both
spectra, as shown in Figure~7.  The MgII broad emission line is at $z = 1.3077 \pm 0.0008$
for the brighter component and $z = 1.3174 \pm 0.0018$ for the fainter
component, overlapping at the 5$\sigma$ level.  A similar agreement
is seen for the CIII] broad emission line.  The narrow-line absorption
  features present in the spectra of both components can be modeled
  with two absorbers at $z = 1.1525 \pm 0.0002$ and $z = 0.7644 \pm
  0.0004$.  Both are anchored by an appropriately redshifted MgII
  $\lambda\lambda$ 2796,2803 doublet and accompanying FeII
  $\lambda\lambda$2382, 2600 absorption lines.  We also identify FeII
  $\lambda\lambda$2344, 2374 and 2586 for the $z = 1.1525$ absorber.
  The flux ratio between the components is also remarkably constant at
  about 2.5:1 (0.4 dex in Figure~7) over the entire IMACS spectral
  range.  
  
Based on the Magellan data presented here, WISE~2329-1258 would be
classified as a lensless twin system.  But Treu et al. (2017,
private communication) have observed it using NIRC2 on Keck-II
behind adaptive optics and detected the lensing galaxy as well as
the lensed quasar host galaxy, so we count it as a confirmed lens.
Like HE 1104-1805, this system also has strong absorption line systems,
and we would not be surprised if one of these turns out to be the lens
redshift. 

\subsection{WISE~2344-3056: a quad at $z=1.30$}

The slit used to obtain a spectrum for WISE~2344-3056 ran along
the line connecting the ellipses in Figure~1, with light from
all four images overlapping on roughly six spatial pixels.
We present only the combined
spectrum in Figure~7.  The quasar redshift is $z = 1.2978 \pm 0.0003$
based on a Gaussian fit to the MgII broad emission line.  The CIII]
  broad emission line is also present at a much lower signal to noise.
  There is at least one intervening absorption system at $z = 0.9472
  \pm 0.0012$ anchored by the MgII $\lambda\lambda$2796,2803
  absorption doublet, several FeII lines
  ($\lambda\lambda$2344,2374,2382,2399,2586,2600), FeIII
  $\lambda$2419, and MgI $\lambda$2852. 
\section{Lens models}

We used Keeton's {\tt lensmodel} program (2001) to fit models
to our astrometry for WISE 2304-2214 and WISE 2344-3056 given in
Tables 2 and 3.

\bigskip
\subsection{WISE 2304-2214}

For WISE 2304-2214 we used a singular isothermal ellipsoid.
The generic expectation is that the quasar images will be micro-lensed
(Witt, Mao and Schechter 1995) and we are reluctant to use their continuum
flux ratos as constraints.  But as this leaves the model underconstrained,
we took the emission line flux ratio to be approximately unity and
adopted this as a constraint.

The model puts the source $0\farcs17$ west and $1\farcs08$ north of
image A and has a lens strength of $1\farcs04$.  The model
magnifications are, -2.6 and 2.6 for images A and B respectively.  By
contrast image A is 0.94 mag brighter than B in the $r$ and $i$
filters.  We take this to be an indication of micro-lensing of the
continuum.  Image A is more likely to be micro-lensed than image B
both because A is a saddle point (Schechter and Wambsganss 2002) and
because B is more than twice as far from the lensing galaxy and
passes through a lower surface density of stars.

The model ellipticty is 0.22 and is directed along P.A. $73\fdg0$ east
of north. as compared to the observed ellipticity and orientation of
0.58 and $111\fdg4$.  Shear of 0.1 from a galaxy $5\arcsec$ toward
P.A.  $30\arcdeg$ from image B produces model with larger
ellipticities that are more nearly aligned with the observation.

\subsection{WISE 2344-3056}

For WISE 2344-3056 we used a singular isothermal sphere with external
shear to fit astrometric data in Table 2.  As we do not detect the
lensing galaxy, the center of the lens is left free and found to be
$0\farcs436$ west and $0\farcs151$ north of image A.  The model puts
the source $0\farcs441$ west and $0\farcs153$ north of image A.  The
lens strength -- which would be the radius of the Einstein ring were
there no shear -- is $0\farcs99$.  The external shear is 0.063 and is
directed along P.A. $-71\fdg5$.  The signed magnifications for images
A, B, C and D are, respectively, -6.8, 8.4, 8.6, -8.1.

\section{Discussion}

\subsection{The method}

We have searched for gravitationally lensed quasars by analyzing
VST-ATLAS image cutouts of red WISE sources.  Those that could be
consistently split into two nearly pointlike objects with roughly
constant flux ratios across multiple filters and quasar-like colors
in those filters (specifically $u - g < -0.5$) were selected as
candidates for followup spectroscopy and imaging.  Three new lenses,
two pairs of lensless twins, one binary quasar and two projected quasar pairs
were found.  Owing to bad weather during 2015 and 2016, only a fraction of
the candidates (albeit some of the best) were observed.

The list of candidates is likely to increase considerably with a) 
completion of the ATLAS survey, b) application of the method to fainter
WISE sources and c) incorporation of a more sophisticated scheme for
gauging whether the $ugriz$ colors of a particular pair of objects are
quasar-like.  The method could be extended to splitting sources into
triples, with a primary goal of more readily identifying quadruply
lensed quasars.

We encountered an unanticipated bottleneck in the speed with which the
OmegaCam Science Archive can produce cutouts -- something on the order
of one cutout per second, far more than one might think necessary for
retrieving 2500 pixels.  The servers for the DES and KiDS surveys are
not qualitatively faster.  This casts a pall on programs that might
require $10^7$ or $10^8$ cutouts.  While it might be difficult to
retool existing archives to speed up the process, we imagine that
future systems will deal with such programs more efficiently.

Our method (modified for use with $grizY$ photometry) is one
of several now being used to search for lenses in the Dark Energy
Survey (Agnello et al 2015; Ostrovski et al 2017), which ought
to produce at least as many lensed quasars as ATLAS.  As no
one method will be perfect, comparison of the results will shed
light on their relative strengths and weaknesses.

\subsection{Lensless Twins}

The VST-ATLAS survey has better seeing than SDSS, allowing, in
principle, the discovery of less widely separated lensed quasars.  But
the galaxies that produce close pairs are fainter than those that
produce wide pairs and are more crowded by the lensed quasar images,
making it more difficult to identify them.

We have argued that it is premature to classify either of our
``lensless twin'' quasars as binary quasars -- fraternal twins --
despite the fact that no lensing galaxy has as yet been identified.
In Figure~8 we show the two systems, with the component spectra
shifted to overlap.  For WISE 0326-3122 the agreement is nearly
perfect, while for WISE 1051-1142 the differences are consistent with
what one would expect for a micro-lensed system.  We note that WISE
2329-1258 was at first classified as a lensless twin system, but in
the time since the original submission of this paper has been
reclassified as lensed quasar with the detection of a lensing galaxy
by other investigators.

With sufficiently deep direct images in sufficiently good seeing one
can set upper limits on the lensing galaxy that rule out the lensing
hypothesis, but only with extensive modeling of lensing scenarios.  If
higher signal-to-noise spectra, or spectra at other wavelengths were
to show significant differences in the line profiles, they would again
rule out the lensing hypothesis.  Confirmation of the lensing
hypothesis might come either from identification of the lensing galaxy
or from correlated variations in the fluxes, as might be obtained from
synoptic observations with the LSST.

\bigskip
\section{Summary}

We have analyzed images from the VST ATLAS survey to identify
candidate gravitationally lensed quasar systems in a sample of WISE
sources with $W1 - W2 > 0.7$.  Results from followup spectroscopy with
the Baade 6.5 m telscope are presented for eight systems.  One of
these is a quadruply lensed quasar, and two are doubly lensed systems.
Two are projected superpositions of quasars at two different
redshifts, and one appears to be a pair of distinct quasars at the
same redshift.  In two systems the component spectra are consistent
with the lensing hypothesis, after allowing for micro-lensing.  But as
no lensing galaxy is detected in these two, we classify them as
lensless twins.  More extensive observations are needed to establish
whether they are lensed quasars or physical binaries.

\clearpage
\begin{figure}
\includegraphics[angle=0,scale=0.80]{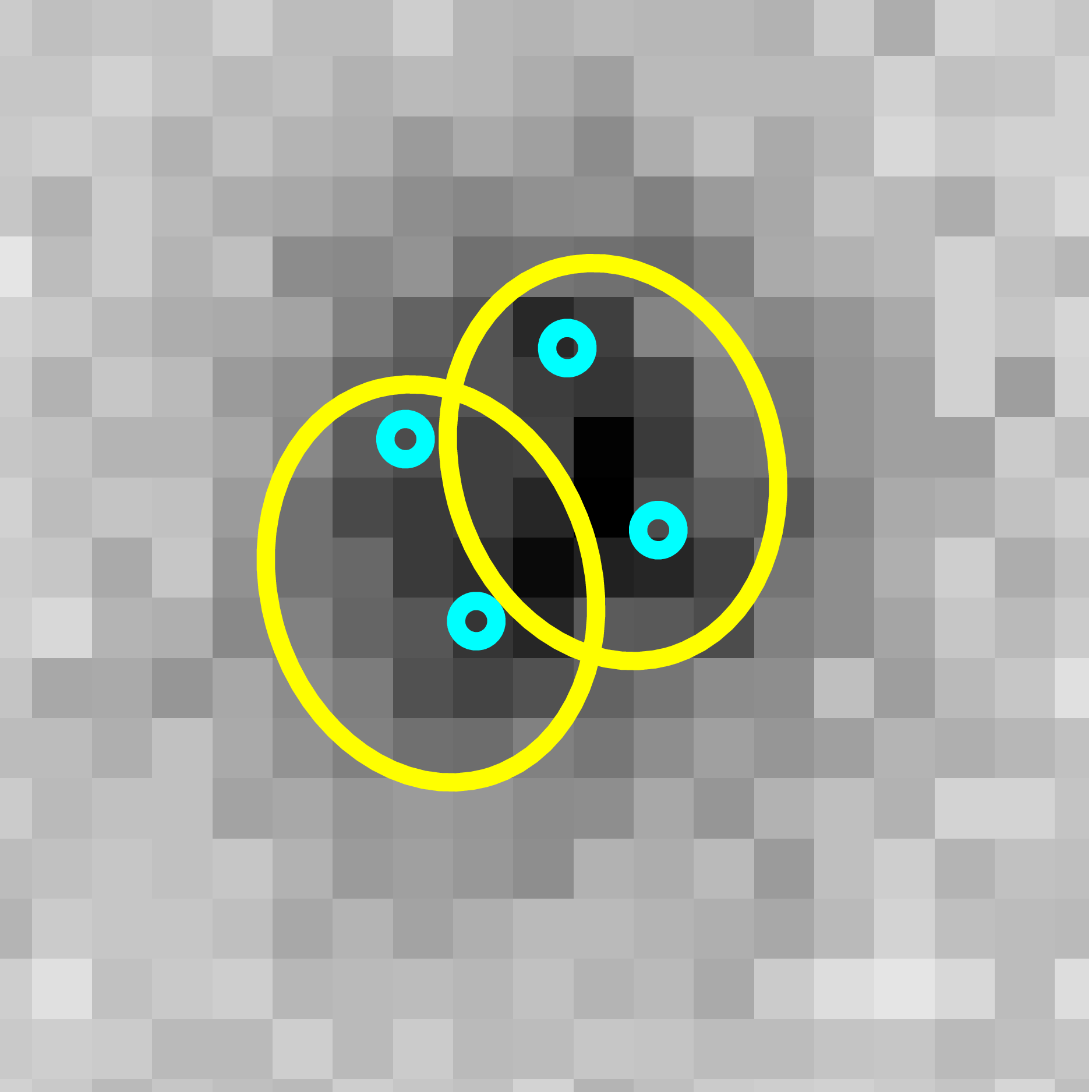}
\caption{The VST-ATLAS $g$ stack for WISE 2344-3056.
The yellow ellipses are the result of our anchor
splitting of this source.  The blue circles are at the positions of
the four images (D, C, B and A from left to right) identified with
IMACS. The scale is $0\farcs213$ per pixel.  North is up and East is
to the right.}
\end{figure}

\clearpage

\begin{figure}
\includegraphics[angle=0,scale=0.90] {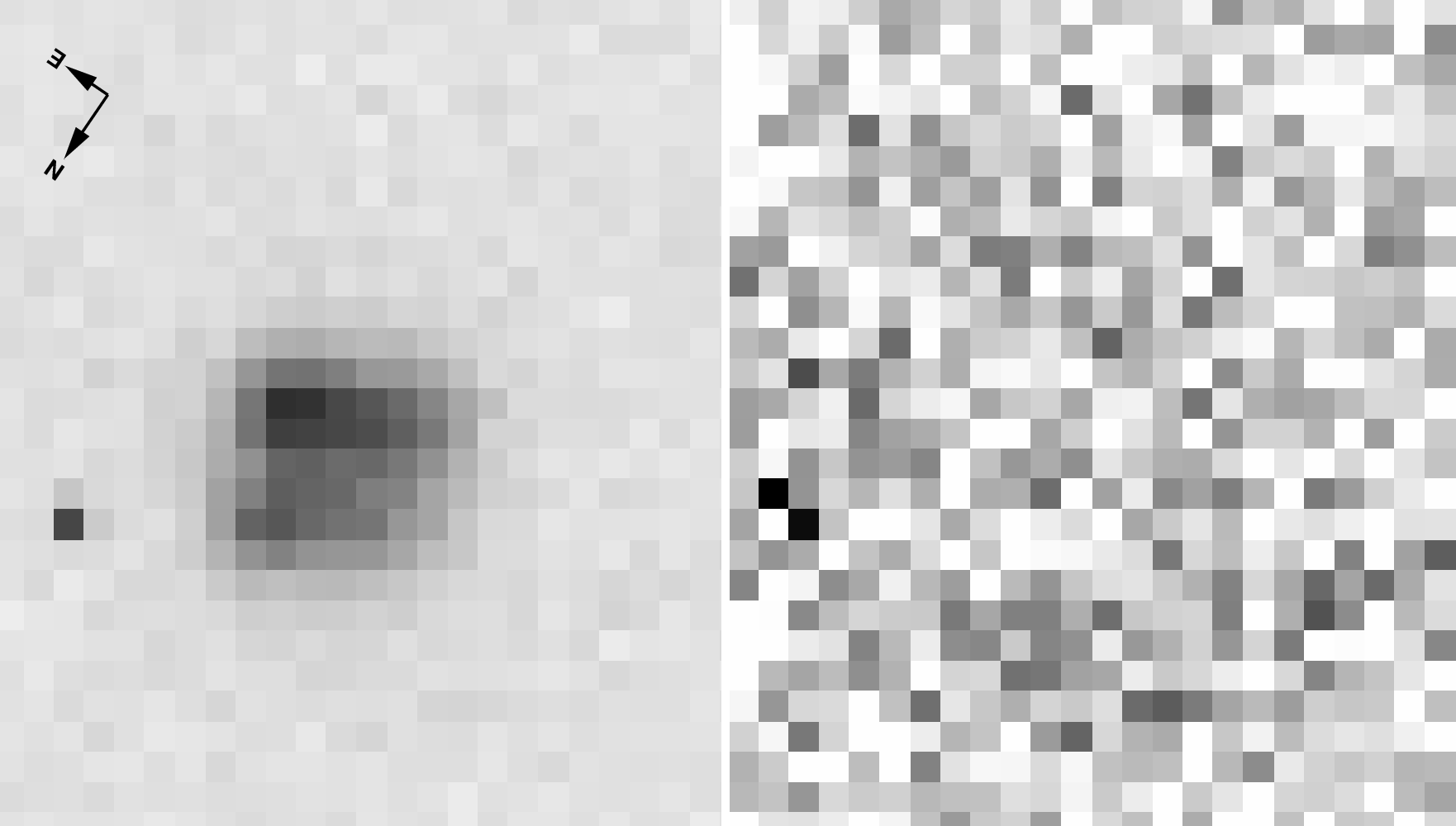}
\caption{Left: A 60s $i$ exposure of WISE 2344-3056 taken with IMACS
in $0\farcs55$ seeing.  
Right: the same exposure, with four point sources
subtracted, at 10 times higher contrast.  The scale is $0\farcs200$ per pixel}
\end{figure}

\clearpage

\begin{figure}
\includegraphics[angle=0,scale=0.90] {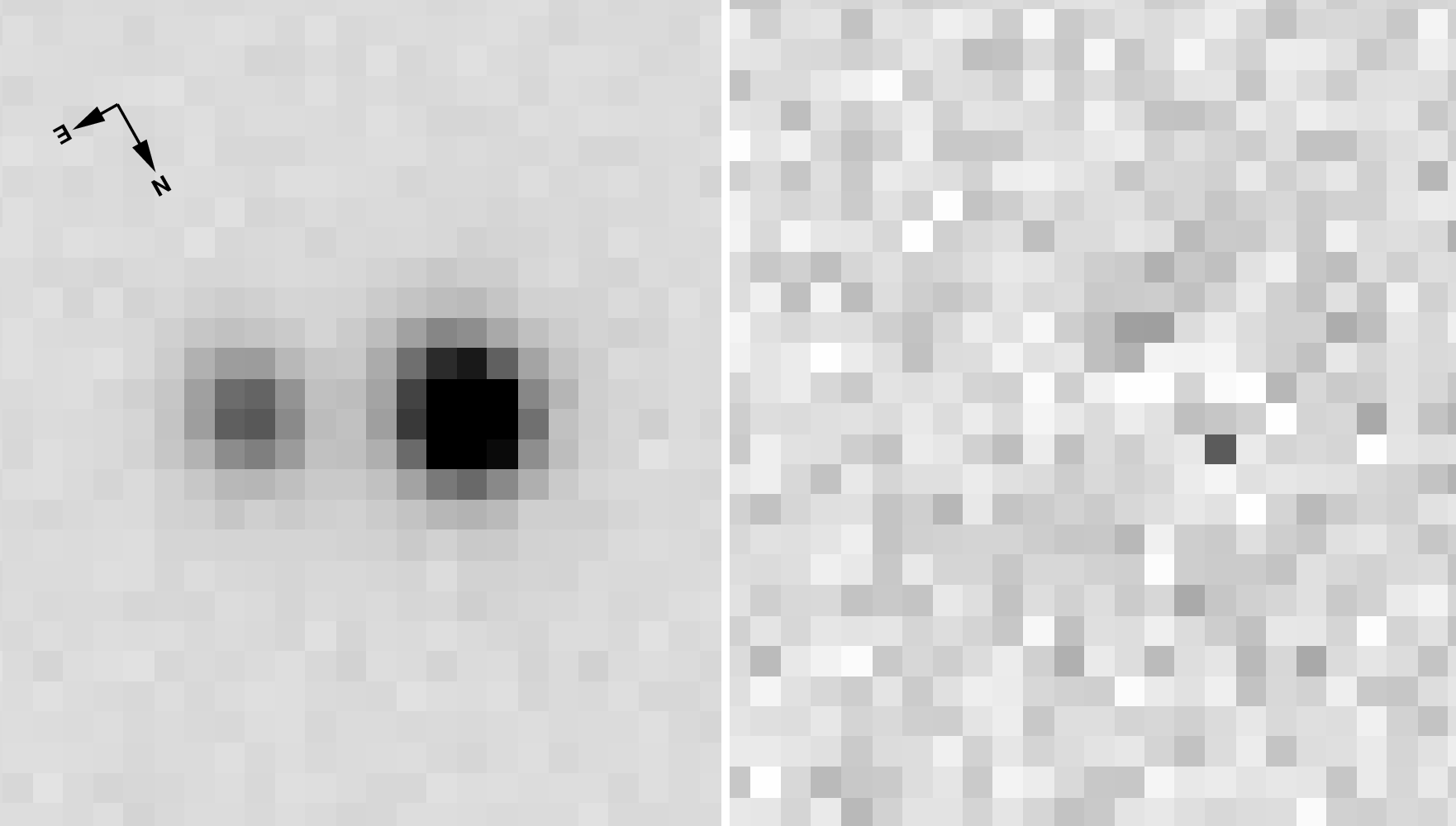}
\caption{Left: A 30s Sloan $r$ exposure of WISE 0326-3122
in $0\farcs64$ seeing.  Right: the same exposure, with two point
sources subtracted, at 5 times higher contrast.  The scale is
$0\farcs200$ per pixel.}
\end{figure}

\clearpage

\begin{figure}
\includegraphics[angle=0,scale=0.90] {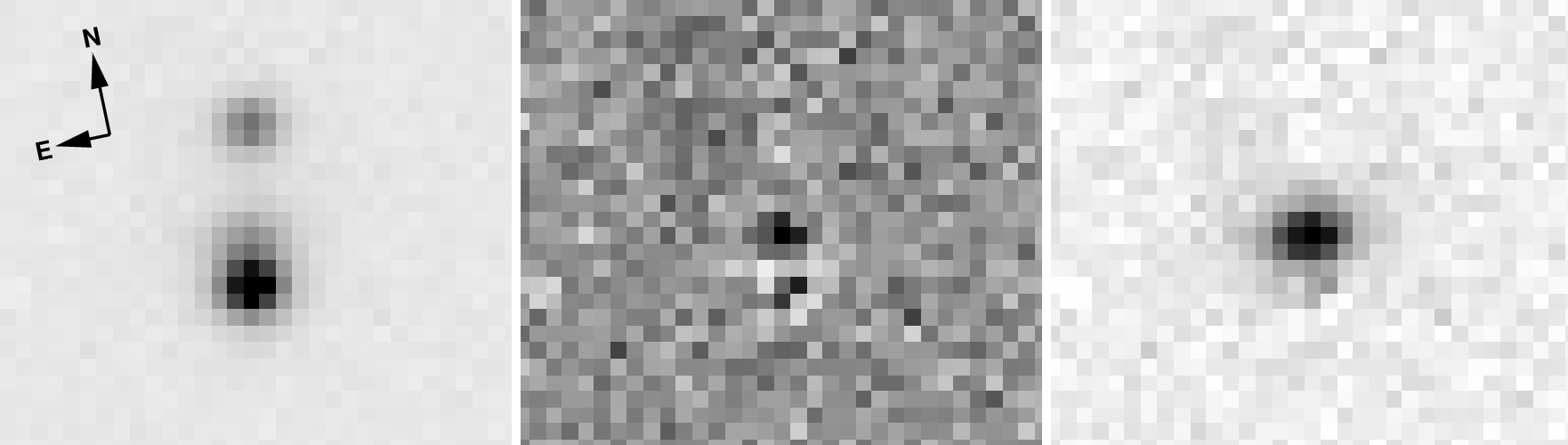}
\caption{Left: A 60s Sloan $i$ exposure of WISE 2304-2214
in $0\farcs64$ seeing.  Center: the same exposure, with two point sources
and an extended source subtracted, at 10 times higher contrast.
Right: the same exposure with only the
point sources subtracted, at 4 times higher contrast.  The scale is
$0\farcs221$ per pixel.}
\end{figure}

\clearpage
\begin{figure}
\includegraphics[angle=0,scale=0.90]{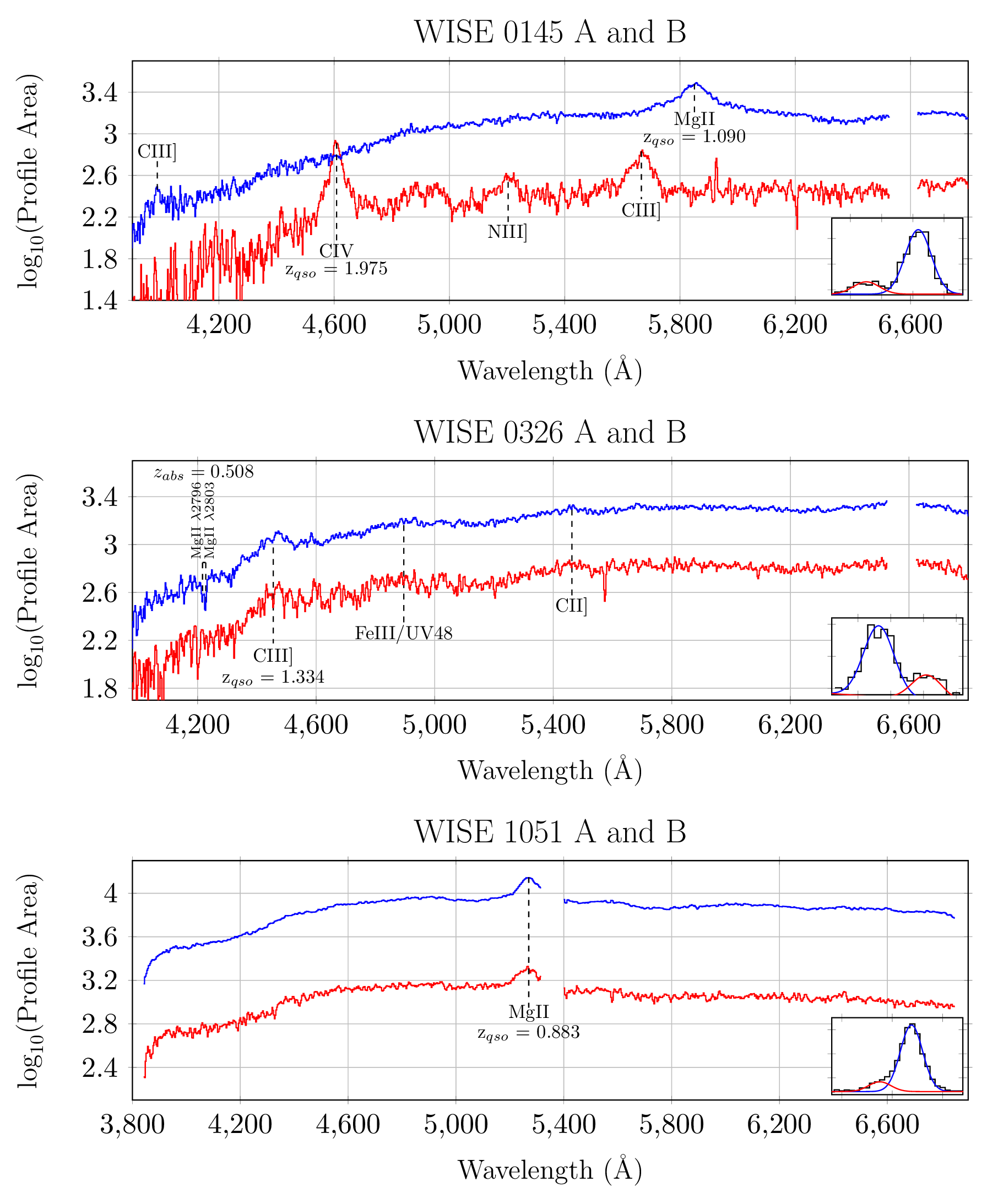}
\caption{IMACS spectra of WISE 0145-1327, WISE 0326-3122,
and WISE 1051-1142.
Image A is brighter than  image B.}
\end{figure}

\clearpage
\begin{figure}
\includegraphics[angle=0,scale=0.90]{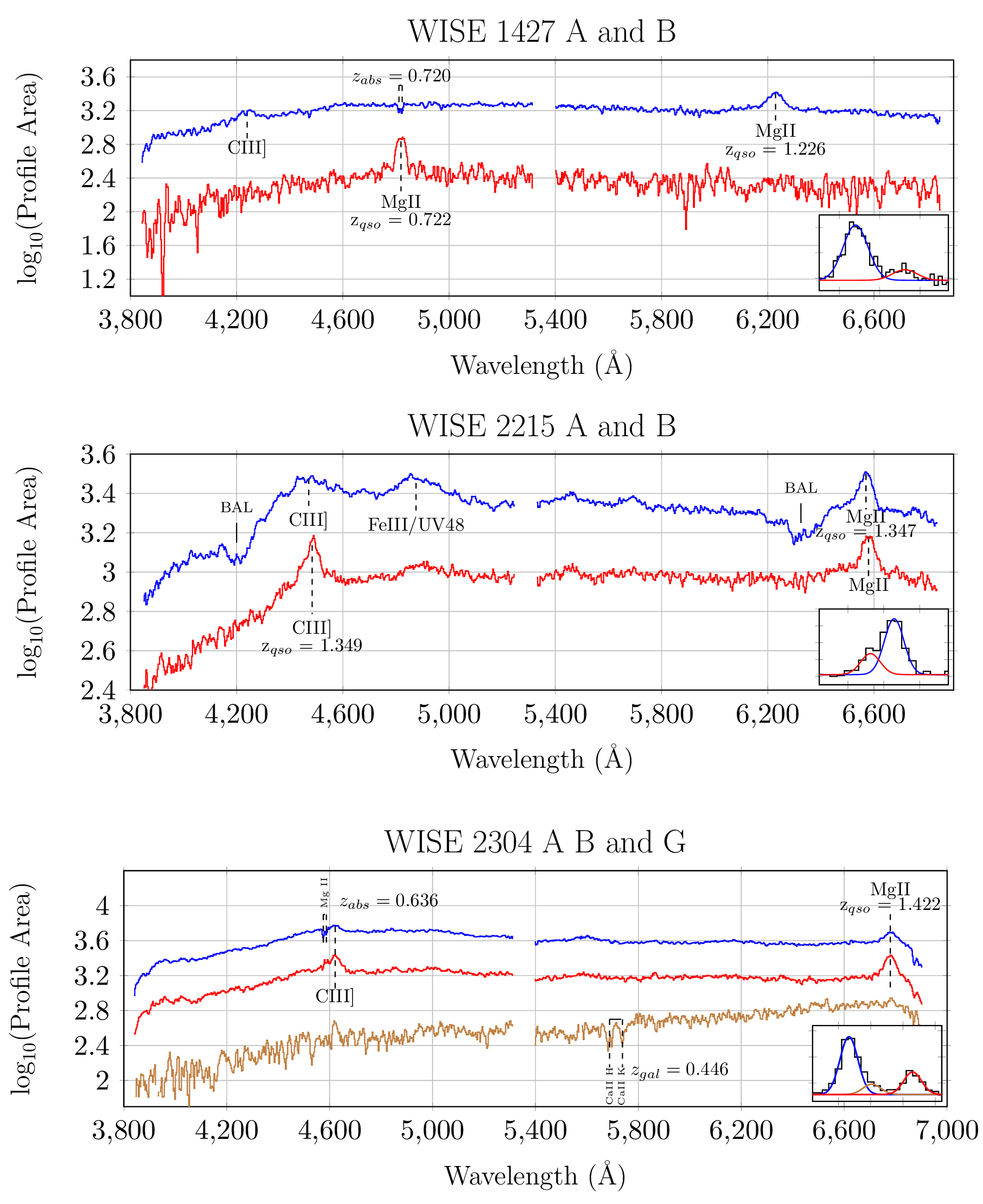}
\caption{IMACS spectra of WISE 1427-0715
and WISE 2215-3056 and WISE 2304-2214.
Image A is brighter than  image B.}
\end{figure}

\clearpage
\begin{figure}
\includegraphics[angle=0,scale=1.20]{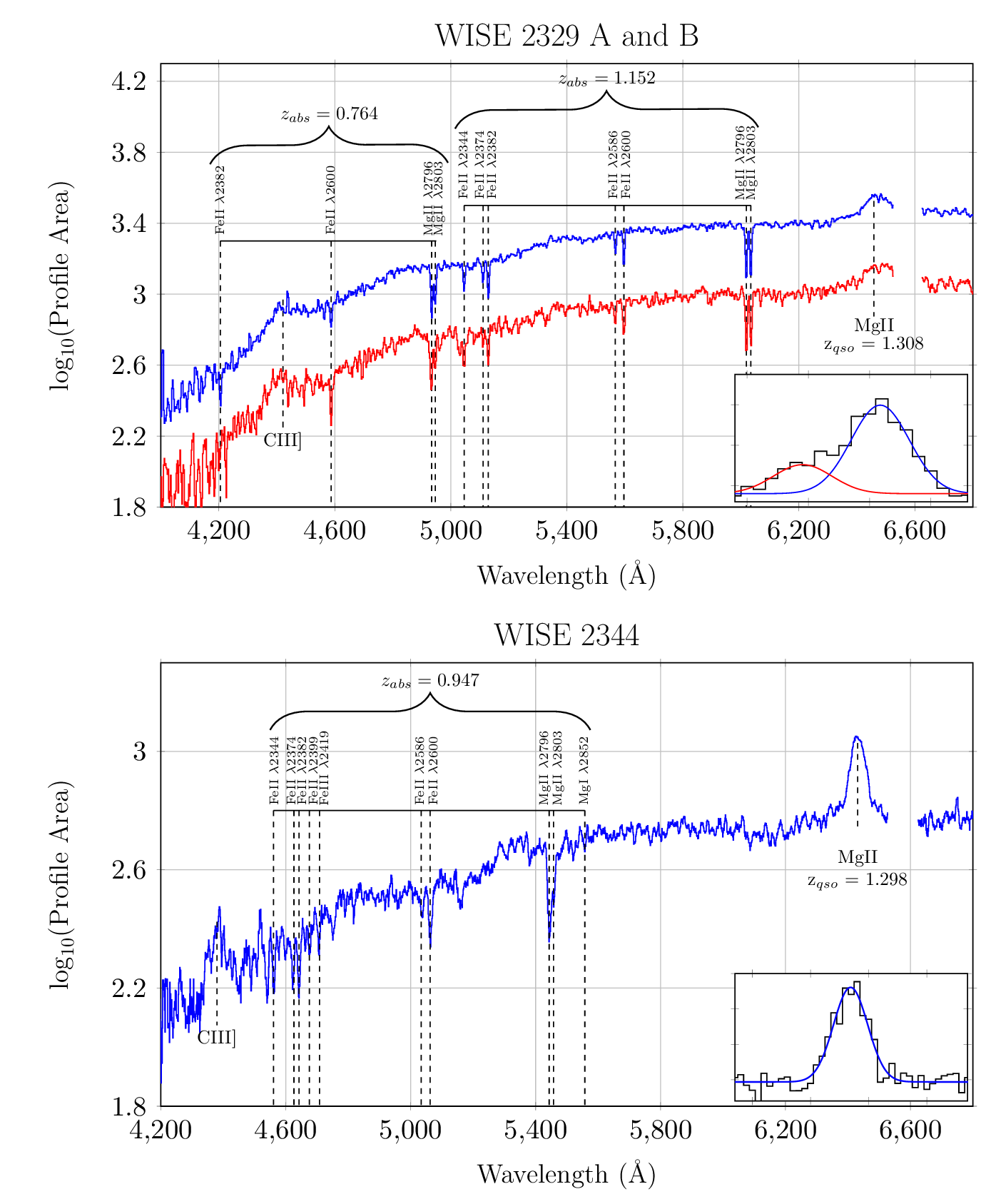}
\caption{IMACS spectra of WISE 2329-1258 A (upper) and B (lower)
and WISE 2344-3056, taken with IMACS}
\end{figure}

\clearpage
\begin{figure}
\includegraphics[angle=0,scale=0.90]{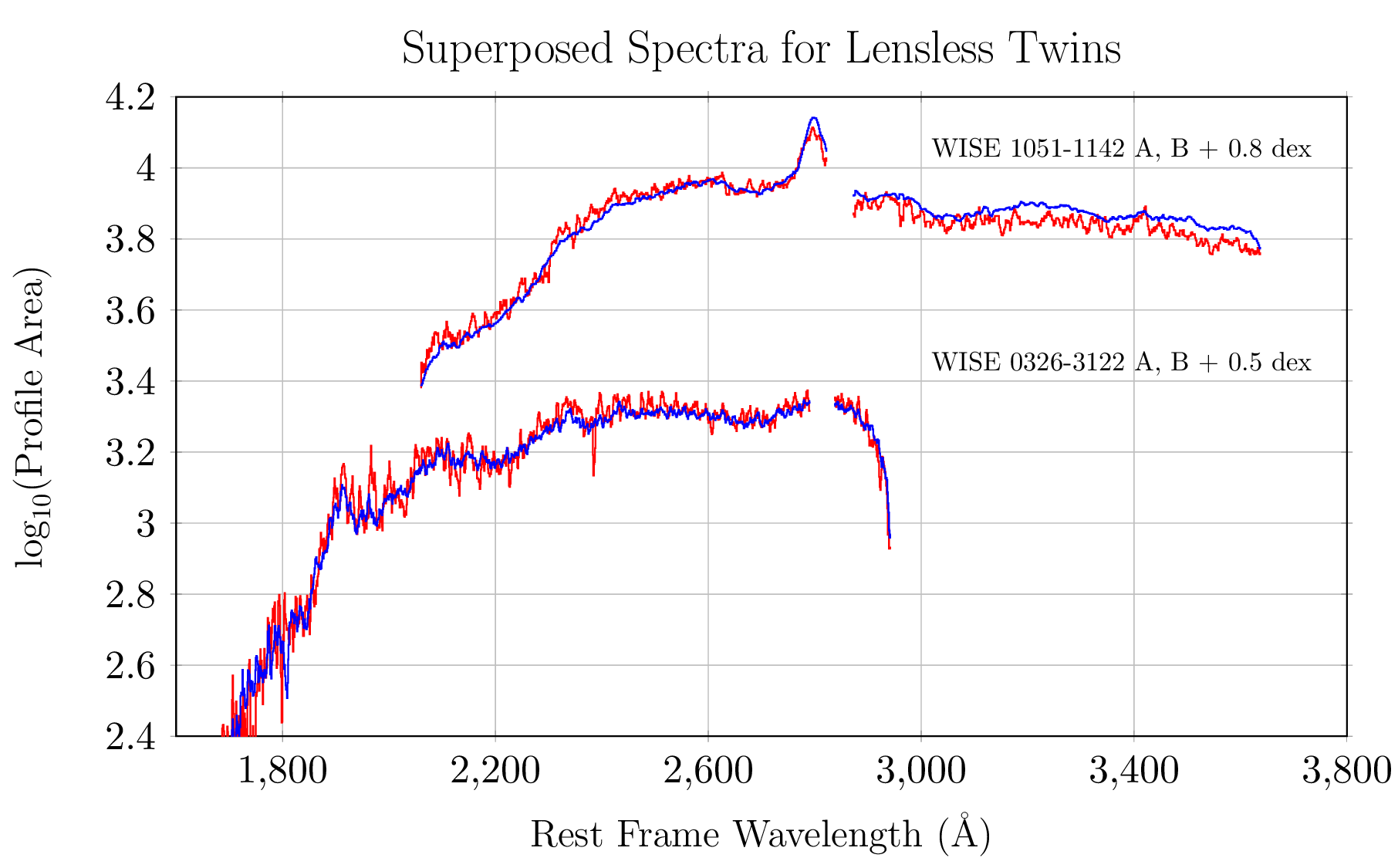}
\caption{Rest frame spectra of WISE 0326-3122 and WISE 1051-1142
with the $A$ and $B$ components superposed.}
\end{figure}

\clearpage
\begin{deluxetable}{llrlrrl}
\tablecolumns{7}
\tablewidth{0pc}
\tablecaption{Spectroscopic Observations of Lensed Quasar Candidates}
\tablehead{\colhead{name} &
  \colhead{R.A.}    & \colhead{$\Delta\theta$} & \colhead{$i_A^*$}    &
  \colhead{score}  & \colhead{camera}         & \colhead{description} \\
&  \colhead{Dec.}    & \colhead{P.A.}           & \colhead{$i_B^*$}  &
  \colhead{$(u - g)^\dagger $}  & \colhead{exp.}   & \colhead{redshift(A/B)}       \\
}
\startdata
  
WISE 0145-1327 &  01 45 25.3  & $1\farcs68$ & 19.25   & 0.39 & f/2 & projected      \\
& $-$13 27 25  & $5^\circ$   & 20.59  & -1.01 & 1200s    & 1.09/1.97 \\
\hline
WISE 0326-3122 & 03 26 06.8  & $1\farcs43$ & 19.49  &    0.60 & f/2 & lensless twins   \\
& $-$31 22 54    & $-60^\circ$   & 20.56 &-1.17 & 1200s & 1.34  \\
\hline
WISE 1051-1142 &  10 51 41.9 & $1\farcs47$    & 17.25  & 0.50 & f/4 & lensless twins   \\
& $-$11 42 39  & $-5^\circ$    & 19.39  &-0.89 & 900s & 0.88    \\
\hline
WISE 1427-0715 &  14 27 04.8 & $2\farcs80$ & 18.92       & 0.49  & f/4 & projected   \\
& $-$07 15 56  & $9^\circ$     &  20.32    & -0.73 & 900s    & 1.23/0.72 \\
\hline  
WISE 2215-3056 &  22 15 25.6  & $0\farcs71$ & 18.31     & 0.29 & f/4 & binary     \\
& $-$30 56 35    & $-41^\circ$   & 19.07 &-0.61  & 900s   & 1.34 \\
\hline  
WISE 2304-2214 &  23 04 25.3  & $2\farcs19$ & 19.57     & 0.58 & f/4 & lensed     \\
& $-$22 14 46    & $-11^\circ$   & 20.50 &-1.37  & 1800s   & 1.42 \\
\hline  
WISE 2329-1258 &  23 29 57.9 & $1\farcs27$  & 17.63  & 0.96 & f/2 & lensed   \\
& $-$12 58 59  & $46^\circ$    & 18.60 & -0.82 & 600s & 1.314 \\
\hline
WISE 2344-3056 &  23 44 17.0 & $2\farcs18$  & 20.31$\ddagger$  & 0.44 & f/2 & quadruple   \\
& $-$30 56 26  & $-12^\circ$   & 20.63$\ddagger$  &-0.63 & 900s & 1.298    \\
\enddata
\tablenotetext{*}{magnitudes for $A$ and $B$ components derived
  from catalogued ATLAS Petrosian AB magnitudes and constrained separation
  flux ratios,.
}
\tablenotetext{\dagger}{$u-g$ colors are in the Vega-like system of
  the ATLAS FITS headers}
\tablenotetext{\ddagger}{$i_C = 20.71$ and $i_D = 21.12$}
\end{deluxetable}

\begin{deluxetable}{lrrr}
\tablecolumns{4}
\tablewidth{0pt}
\tablecaption{Astrometry\tablenotemark{*} and Photometry for WISE 2344-3056}
\tablehead{
\colhead{image}  & \colhead{$\Delta \alpha$} & \colhead{$\Delta \delta$} & $i$\\
}
\startdata
A &  0\farcs000 & 0\farcs000  & 20.31 \\
B & -0\farcs304 & 0\farcs665  & 20.63 \\
C & -0\farcs641 &-0\farcs337  & 20.71 \\
D & -0\farcs886 & 0\farcs293  & 21.12 \\
\enddata
\tablenotetext{*}{Positions in arcseconds relative to 
  $\alpha_A =   23^h44^m16\farcs995$ and
  $\delta_A = -30^\circ56\arcmin26\farcs22$}
\end{deluxetable}

\begin{deluxetable}{lrrrr}
\tablecolumns{5}
\tablewidth{0pt}
\tablecaption{Astrometry\tablenotemark{*} and Photometry\tablenotemark{\dagger} for WISE 2304-2214}
\tablehead{
  \colhead{image}  & \colhead{$\Delta \alpha$} & \colhead{$\Delta \delta$} &
  \colhead{$r$} & \colhead{$i$}\\
}
\startdata
A &  0\farcs000 & 0\farcs000  & 19.52 & 19.57 \\
B & -0\farcs431 & 2\farcs145  & 20.46 & 20.50 \\
G & -0\farcs093 & 0\farcs673  & 21.48 & 20.65 \\
\enddata
\tablenotetext{*}{Positions in arcseconds relative to 
  $\alpha_A =   23^h04^m25\farcs348$ and
  $\delta_A = -22^\circ14\arcmin46\farcs95$}
\tablenotetext{\dagger}{magnitudes 
  from three component fits to Magellan images with scale
  set by ATLAS AB magnitudes for template stars}
\end{deluxetable}

\end{document}